\documentstyle[aps,eqsecnum,epsfig]{revtex}

\def\Journal#1#2#3#4{{#1} {\bf #2}, #3 (#4)}

\def\NPB{{\em Nucl. Phys.} B}
\def\NPPS{{\em Nucl. Phys.} B {\em Proc. Suppl.}}
\def\PLB{{\em Phys. Lett.}  B}
\def\PRL{\em Phys. Rev. Lett.}
\def\PRD{{\em Phys. Rev.} D}
\def\PRe{\em Phys. Rep.}

\def\AnP{\em Ann. Phys. (NY)}
\def\JETP{\em JETP Lett.}

\def\beq{\begin{equation}}
\def\eeq{\end{equation}}

\def\ts{\left(}
\def\td{\right)}
\def\qs{\left[}
\def\qd{\right]}
\def\esp{\mbox{e}}
\def\ag{\alpha}

\def\bg{\beta}

\def\dg{\delta}

\def\lg{\lambda}

\def\sg{\sigma}

\def\Dg{\Delta}

\def\Og{\Omega}
\def\Phu{\Phi^{(1)}}
\def\Phd{\Phi^{(2)}}
\def\Phux{\Phu (x)}
\def\Phdx{\Phd (x)}
\def\Phuxp{\Phu\, ' (x)}
\def\Phdxp{\Phd\, ' (x)}
\def\Phdxs{\Phd\, '' (x)}

\def\phu{\phi^{(1)}}
\def\phd{\phi^{(2)}}
\def\phux{\phu (x)}
\def\phdx{\phd (x)}

\def\phdxs{\phd\, '' (x)}
\def\Tr{\mbox{Tr}}
\def\esp{\mbox{e}}

\begin{document}
\setlength{\baselineskip}{20pt}
\renewcommand{\arraystretch}{2}
\jot=20pt
\abovedisplayskip=20pt
\belowdisplayskip=20pt
\abovedisplayshortskip=20pt
\belowdisplayshortskip=20pt

\title{\bf Center Vortices and Monopoles without lattice Gribov copies}

\author{Ph. de Forcrand}

\address{Inst. f\"ur Theoretische Physik, ETH H\"onggerberg, 
CH-8093 Z\"urich, Switzerland \\
and \\
CERN, Theory Division, CH-1211 Gen\`eve 23, Switzerland \\
E-mail: forcrand@itp.phys.ethz.ch} 

\author{M. Pepe}

\address{Inst. f\"ur Theoretische Physik, ETH H\"onggerberg, 
CH-8093 Z\"urich, Switzerland \\
E-mail: pepe@itp.phys.ethz.ch} 

\maketitle

\begin{abstract}
We construct a smooth gauge for the adjoint field which is free of
ambiguities on the lattice. In this Laplacian Center Gauge, center
vortices and monopoles appear together as local gauge defects.
A numerical study of center vortices in $SU(2)$ and $SU(3)$ supports
equality of the $Z_N$ and $SU(N)$ string tensions in the continuum
limit, and only then.
\end{abstract}


\section{Introduction}

Ever since the advent of QCD, one has tried to isolate a subset
of collective degrees of freedom (d.o.f.) which would be responsible for the
non-perturbative features of the $SU(2)$ or $SU(N)$ Yang-Mills theory, 
in particular for confinement.
It is natural to associate non-perturbative effects with topological 
excitations. Therefore, all the non-trivial homotopy groups 
$\Pi_3(SU(2))={\bf Z}$, $\Pi_2(SU(2)/U(1))={\bf Z}$ and $\Pi_1(SU(2)/Z_2)=Z_2$, 
and their $SU(N)$ homologues have been considered. 
The corresponding topological 
excitations have codimension 4, 3 and 2 respectively, and are instantons,
Abelian monopoles and center vortices. Among these, the latter have been
emerging as particularly worthy of attention. Center vortices correspond
to non-contractible Wilson loops in the adjoint representation. The simplest
example is obtained with an $SU(2)$ gauge field along a Wilson loop of length
$L$: $A_\mu^1 = A_\mu^2 = 0, A_\mu^3(x) = 2\pi / L$.
The Wilson loop $exp(i \oint_L dx_\mu A_\mu^a(x) \frac{\sigma^a}{2}) = exp(i \pi \sigma_3)$ has
trace $exp(2 i \pi \sigma_3) = 1$ in the adjoint representation, but is
not contractible to the identity. 

Numerical lattice simulations have given evidence for the relevant
role played by center vortices.
Now, topology is a feature of continuum field theory,
and studying topological excitations on the lattice requires some kind of 
interpolation. Non-contractible adjoint loops are rather elusive under 
direct investigation, so that an indirect approach has been successfully
developed \cite{green1}, in analogy with the detection of Abelian monopoles.
Any topological excitation can be exposed as a local gauge-fixing singularity
in an otherwise smooth gauge. Abelian monopoles appear as singularities
in the Maximal Abelian Gauge, which minimizes the non-Abelian components
$\int d^4x \sum_\mu(|A_\mu^1(x)|^2 + |A_\mu^2(x)|^2)$.
Similarly here, gauge-fixing is used to bring each gauge link as close as possible
to a center element, hence the name Maximal Center Gauge. This is equivalent
to bringing all adjoint links close to the identity, that is, making the
adjoint field as smooth as possible.
Each fundamental link matrix $U$ is then decomposed into a part 
$U' \in SU(N)/Z_N$,
which reproduces the smooth adjoint field and in addition remains close to the $SU(N)$
identity, and a $z \in Z_N$ part. The latter seems at first sight irrelevant for
the detection of non-contractible adjoint loops. However, such a loop 
as in the $exp(i \pi \sigma_3)$ example above, will have a trace in the
fundamental representation equal to $-1$, or more generally to a non-trivial
$Z_N$ element. 
If one neglects the fluctuations of the $U'$ part, center
vortices should appear as non-trivial $Z_N$ excitations, localized on 
plaquettes and called $P$-vortices. $P$-vortices form closed co-dimension 2
surfaces in the dual space--time, which disorder the $Z_N$ ensemble $\{z\}$.
They are localized defects, associated with macroscopic center vortices
via gauge-fixing.
This indirect strategy, called center projection, has been used to assess
numerically the importance of center vortices. Numerical evidence has been 
accumulating, which shows that this $P$-vortex disorder
generates a $Z_N$ string tension close in value to the original $SU(N)$ one
\cite{green1}. A physical density of vortices per ${\rm fm}^2$ can be extracted
\cite{green2}. $SU(N)$ chiral symmetry breaking induces a quark condensate and
Dirac zeromodes in the $Z_N$ ensemble also \cite{phmax2}. 
On the other hand, the coset $\{U'\}$ ensemble shows no confinement, 
no chiral symmetry breaking, and no topological charge \cite{phmax1}. 
While this may be a simple kinematic effect,
it is tempting to adhere to the idea of center dominance, according
to which center vortices, and nothing else, are responsible for the 
non-perturbative Yang-Mills features. One implication of center dominance is 
that the string tension produced by center vortices should match the Yang-Mills
string tension exactly.

However, this construction rests on shaky ground. The connection between $Z_N$
excitations which are located on a plaquette and their $SU(N)$ ancestors,
center vortices believed to be of macroscopic size 
-- smooth examples of which have been explicitly constructed \cite{TGA} -- 
proceeds via gauge fixing.
When each adjoint link is maximally close to the identity, the assignment
of a $Z_N$ center element to the fundamental link, i.e. the center projection
itself, can be performed with the most confidence.
However, the technical difficulty of fixing a smooth gauge on the lattice is
well known. Gauge fixing proceeds by iterative local maximization of a gauge
functional, and this procedure terminates when a local maximum is reached,
without any guarantee, or in fact any reasonable hope, of reaching the global
maximum. Which local maximum is reached depends on the starting point along
the gauge orbit. The various gauges reached correspond to ``lattice Gribov copies''.
This technical problem, which plagues the Landau and Coulomb gauges on the lattice,
is commonly believed to be rather harmless there. In Maximal Center Gauge,
its harmful effects have been exhibited \cite{K&T}: the copy obtained when
starting the gauge-fixing procedure from Landau gauge corresponds to a very high
value of the gauge-fixing functional, higher than a typical local maximum,
but gives, after center projection, $Z_N$ excitations which do not confine.
Evidence has been produced that, the higher the value of the local maximum one
reaches, the smaller the $Z_N$ string tension \cite{borny}. Although this
evidence is currently under dispute \cite{bertle}, the technical problem
of lattice Gribov copies prevents a reasonable degree of confidence in the
idea of center dominance.

The present paper, we believe, should restore full confidence in this idea.
Here, we address and solve the problem of the gauge ambiguity.
This problem was already solved for the Landau gauge in Ref.\cite{vink}.
There, a different gauge was proposed, the Laplacian gauge, which is 
Lorentz-symmetric and produces 
a smooth gauge field like the Landau gauge, but has no lattice Gribov copies.
The advantages of such an unambiguous gauge have barely been explored
\cite{vink2,Pepe}. In the present paper, we generalize the construction
of \cite{vink} to $SU(N)$ in the adjoint representation, since we want 
the adjoint field to be smooth. In our ``Laplacian Center Gauge'', the 
adjoint $SU(N)/Z_N$ field is uniquely gauge-fixed -- aside from exceptional
configurations which are genuine Gribov copies--, and a remaining local $Z_N$
gauge freedom subsists. Now, even though the adjoint field is uniquely gauge-fixed
in general, there exists a sub-manifold of points where the gauge remains
ill-defined. 
These local gauge ambiguities are the necessary companions of macroscopic 
topological excitations. As explained earlier, a non-contractible adjoint
Wilson loop will prevent the adjoint field from being smooth everywhere;
similarly, an Abelian monopole will correspond to a singularity of the
Abelian projected field. Indeed, by studying the details of the gauge condition,
we show that on some subset of points the gauge freedom is locally enlarged 
from $Z_2$ to $U(1)$, and from $U(1)$ to $SU(2)$, defining two types
of local ambiguities which can be readily identified
with center vortices and Abelian monopoles, the latter being embedded
in the former, as anticipated in \cite{green+1,green+2}. 
These topological objects appear naturally together as 
local gauge defects. This unified description reconciles with center 
dominance the large body of numerical evidence in favor of monopole 
dominance: Abelian monopoles do not represent an alternative to center
vortices as effective degrees of freedom; they are their undissociable partners.
In particular, the condensation of monopoles in the confined phase of the
$SU(N)$ theory \cite{thooft1,mandel} implies the percolation of center
vortex surfaces, by the following simple argument. A non-zero monopole
condensate means that a single monopole can be created at point $(x,t)$,
or equivalently that monopoles are not pair-wise confined. Since monopole 
world--lines are closed loops, the monopole at $(x,t)$ must have a partner
at $(x',t)$, and a ``single'' monopole simply has a partner at infinity.
This in turn means that monopole world--lines percolate. Since we show that
monopole world--lines are embedded in center vortex surfaces, percolation
of one implies percolation of the other.

We complement the qualitative study of local gauge defects outlined above
with a quantitative study, for $SU(2)$ and $SU(3)$. We show that, in 
Laplacian Center Gauge, the center-projected ensemble
confines, while the coset ensemble does not.
This confirms earlier $SU(2)$ results
\cite{green1,phmax1}, and firms up the sketchy $SU(3)$ evidence \cite{SU3},
all obtained in the ambiguous Maximal Center Gauge. But the scenario
of center dominance demands {\em exact} equality of the $Z_N$ and $SU(N)$
string tensions. We test this scenario by considering different Laplacian
Center Gauges, obtained from different lattice discretizations of the 
Laplacian. At finite lattice spacing $a$, different gauges give widely
different $Z_N$ string tensions, which at first sight looks like a 
terrible blow to center dominance. Nevertheless, we show that, in all 
three gauges we consider and presumably in any --unambiguous-- gauge,
equality of the $Z_N$ and $SU(N)$ string tensions is approached as
$a \rightarrow 0$. Therefore, our results support the restoration of center 
dominance in the continuum limit, and only then.

Section II of our paper gives an explicit construction of the Laplacian
Center Gauge for $SU(N)$. Section III discusses the local gauge ambiguities
and their identification as monopoles and center vortices. Section IV
presents our numerical results for $SU(2)$ and $SU(3)$. Section V shows
the effect of alternative gauges and discusses center dominance. 
A final summary and discussion is presented in Section VI.


\section{Laplacian Center Gauge fixing}

The Laplacian gauge was originally proposed by Vink and Wiese \cite{vink}, who
suggested  to use the eigenvectors of the
covariant Laplacian operator to fix the gauge in a non--Abelian gauge
theory: the eigenvectors transform covariantly, and one can fix the gauge 
by prescribing their color orientation at each space-time point.
This method takes advantage of a non--local procedure to fix a 
covariant smooth gauge in an unambiguous way. There are exceptional
cases, genuine Gribov copies, in which the gauge fixed configuration is not unique: however
such cases can be easily detected by the presence of degenerate
eigenvalues and -- from the viewpoint of numerical simulations --
never occur. The perturbative formulation of the Laplacian gauge in $SU(2)$ has
been studied in \cite{PvB}, while the problem of its renormalizability is
still open.
Since we are interested in reducing the symmetry of the gauge
group from $SU(N)$ to its center $Z_N$, it is
useful to consider the Laplacian operator in the adjoint
representation. In fact, since the adjoint representation is invariant
under gauge transformations in $Z_N$, the adjoint Laplacian
procedure fixes unambiguously the gauge up to the center symmetry,
and the Laplacian Center Gauge is just another name for the adjoint Laplacian gauge.
The first step towards this construction was already considered, in $SU(2)$,
by A. van der Sijs \cite{vds}.

Consider 4--dimensional lattice $SU(N)$ gauge
theory. The adjoint Laplacian operator $\Dg^{ab}_{xy} (\dot{U})$ is given by
\beq\label{adjlap}
-\Dg^{ab}_{xy} (\dot{U}) = \sum_{\mu} 
\ts 2\,\dg_{x,y}\,\dg^{ab} - \dot{U}_{\mu}^{ab} (x) \,\dg_{y,x+\hat{\mu}}
- \dot{U}_{\mu}^{ba} (x-\hat{\mu}) \,\dg_{y,x-\hat{\mu}}\td
\eeq
where $a$,$b = 1,\ldots (N^2-1)$ are color indices and $x$,$y$ are
space--time lattice coordinates. The dotted $\dot{U}_{\mu} (x)$
are the link variables in the adjoint representation and are related
to the links $U_{\mu} (x)$ in the fundamental by
\beq\label{adjlink}
\dot{U}_{\mu}^{ab} (x) = \frac{1}{2}\Tr 
\ts \lg _a U_{\mu} (x) \lg _b U_{\mu}^{\dagger} (x) \td
\eeq
$\lg_i$, $i=1,\ldots ,(N^2-1)$ being the generators of $SU(N)$  with the
normalization $\Tr (\lg _a \lg_b)=2\dg_{ab}$. If $V$ is the
volume of the lattice, $\Dg (\dot{U})$ is a $[(N^2-1)V]\times [(N^2-1)V]$ real
symmetric matrix which depends on the gauge field. 
The eigenvalues $\mu_j$ of $\Dg$ are real and the eigenvector equation is
\beq\label{eigeq}
\Dg^{ab}_{xy} (\dot{U}) \phi^{(j)} _b (y) = \mu_j \,\phi^{(j)} _a (x) 
\eeq
where $\phi^{(j)}$, $j=1,\ldots ,[(N^2-1)V]$ are the $[(N^2-1)V]$--dimensional (real)
eigenvectors. So we can associate $(N^2-1)$--dimensional real vectors
$\phi^{(j)} (x)$ to every lattice site. \\
Let us consider a gauge transformation 
$U_{\mu}' (x) = \Og (x) U_{\mu} (x) \Og^{\dagger} (x+\hat{\mu})$ on the
fundamental links. Then, making
use of the property $\dot{(AB)}=\dot{A}\dot{B}$, the eigenvector
equation (\ref{eigeq}) becomes
\beq
\dot{\Og}^{\dagger}\,^{ai} (x) \Dg^{ik}_{xy} (\dot{U}')  \dot{\Og}^{kb} (y)
\phi^{(j)} _b (y) =  \mu_j \,\phi^{(j)} _a (x)
\eeq
This relation shows that the eigenvalues are gauge invariant and the
eigenvectors transform according to 
$\dot{\Og}^{ab} (x) \phi^{(j)} _b (x) = \phi^{(j)}_a \, ' (x) $. This transformation 
law can be rewritten as follows
\beq\label{trlaw}
\Og (x) \Phi^{(j)} (x) \Og^{\dagger} (x) = \Phi^{(j)} \,' (x)
\eeq
where we have defined the $su(N)$ matrices (i.e. in the $SU(N)$ algebra)
$\Phi^{(j)} (x)=\sum_{a=1}^{N^2-1} \phi^{(j)}_a (x) \lg_a$ and
$\Phi^{(j)}\,' (x)=\sum_{a=1}^{N^2-1} \phi^{(j)}_a\,' (x) \lg_a$. 
Gauge transformations rotate the vectors $\phi^{(j)} (x)$ in color space and 
so we can fix the gauge by requiring a conventional arbitrary orientation 
for the $\phi^{(j)} (x)$. This orientation may depend on
the lattice site and it is fixed once and for all; the simplest choice is
to make it space--time independent. We will see below that to perform
the reduction of the gauge symmetry from $SU(N)$ to $Z_N$ we only need 
to fix the orientation of two eigenvectors of the Laplacian
operator. Since we are interested in
fixing a smooth gauge, we
consider the two lowest eigenmodes $\phi^{(1)}$ and $\phi^{(2)}$,
associated with the smallest eigenvalues of the Laplacian. 

The gauge fixing procedure can be split in two steps. In the first, 
one rotates $\Phux$ at every $x$ so that $\Phuxp$ is diagonal, i.e. in the Cartan
subalgebra of $su(N)$. This leaves a residual symmetry
corresponding to gauge transformations belonging to the Cartan
subgroup $U(1)^{N-1}$. 
Therefore, at that stage we already have obtained an unambiguous
Abelian gauge, called Laplacian Abelian Gauge in \cite{vds}.
Nothing more can be done with only one 
eigenvector. To further reduce the gauge freedom we must consider a 
second step where the second eigenvector $\phi^{(2)}$ is taken into
account. The gauge transformation that has rotated $\Phux$ to the Cartan 
subalgebra, maps $\Phdx$ to $\Phdxp$. While $\Phuxp$ is invariant 
under gauge transformations in $U(1)^{N-1}$, this is in general not the
case for $\Phdxp$ (special cases are discussed later). The
gauge symmetry can now be reduced to $Z_N$ by considering gauge transformations
in $U(1)^{N-1}$ which make some conventionally chosen color
components of the twice rotated matrix $\Phdxs$ vanish. The remnant 
center gauge freedom, consistently with the construction, can not be
fixed within the described procedure in the adjoint
representation. In fact, as follows from equation (\ref{trlaw}),
the matrices $\Phi^{(j)} (x)$ are invariant under gauge transformations in the
center group $Z_N$ (these gauge transformations are the identity in
the adjoint representation). Now we describe explicitly how to perform the
presented two--step program. 

{\bf{Step 1}}\\
The starting point is equation (\ref{trlaw}) for the first eigenvector 
$\phu$. $\Phux$ is an $N\times N$ (traceless) hermitian matrix,
and so there exist $N\times N$ unitary matrices $\Og (x)$ which diagonalize it
(we recall that the diagonal elements are the eigenvalues of
$\Phux$). Diagonalizing $\Phux$ is equivalent to requiring that $\Phuxp$ be in the Cartan subalgebra of
$su(N)$. Since we want $\Og (x)$ to be a gauge transformation in $SU(N)$,
we must require that $\det (\Og (x))=1$. The condition that $\Phuxp$ is
diagonal does not specify $\Og (x)$ -- i.e. the gauge symmetry -- up to 
$U(1)^{N-1}$ transformations yet. In fact there are non-diagonal
$SU(N)$ gauge transformations which, when applied to $\Phuxp$, have the only
effect of exchanging the position of the eigenvalues along the
diagonal. In order to eliminate this permutation arbitrariness, it is
necessary to impose an ordering of the eigenvalues.
This is simple to do since the eigenvalues of a hermitian matrix are real.
Once some conventional ordering has been chosen, $\Og (x)$ is really defined up to 
$U(1)^{N-1}$ transformations. 
A last remark concerns the explicit evaluation of $\Og
(x)$, which, in view of (\ref{trlaw}), is the problem of diagonalizing a
hermitian matrix. For the $SU(2)$ and $SU(3)$ gauge groups, this
evaluation can be performed analytically. For larger groups $SU(N\geq 4 )$,
one can simply make use of numerical methods.

{\bf{Step 2}}\\
In {\bf{Step 1}} we have found $\Og (x)$ defined up to gauge
transformations $V(x) \in U(1)^{N-1}$. A second step is necessary 
to reduce the symmetry from $U(1)^{N-1}$ to $Z_N$; this means that we
must introduce a criterion which specifies $V(x)$ up to $Z_N$.
Consider the second eigenvector $\phd$ and apply the gauge transformation
$\Og (x)$ to $\Phdx$ so as to obtain
$\Phdxp = \Og (x) \Phdx \Og^{\dagger} (x)$. Now we look for $V(x)$
such that the vector $\phdxs$, obtained from 
$\Phdxs = V (x) \Phdxp V^{\dagger} (x)$,
has a conventional orientation in color space. Also in this case there is no
need that this orientation be the same at all $x$ but, however, this
is the simplest choice. In order to make the discussion more
transparent, we explicitly consider the $SU(3)$ case; the
generalization to $SU(N)$ is straightforward.\\

A generic matrix $V(x) \in U(1)^{2}$ can be written in the form 
$V(x) = \mbox{diag} \ts\esp^{i(2\ag (x) +\bg (x))}, 
\esp^{i(-\ag (x) +\bg (x))}, \esp^{-i(\ag (x)+2\bg (x))}\td$ with 
$\ag (x)$, $\bg (x)\in [ 0,2\pi )$. We can fix $V(x)$ up to gauge 
transformations in $Z_3$ by imposing the following requirements:
\beq\label{step2req}
\begin{array}{l}
\mbox{1) } \Tr \ts \Phdxs \, \lg_2 \td =0 \\
\mbox{2) } \Tr \ts \Phdxs \, \lg_7 \td =0 \\
\mbox{3) } \Tr \ts \Phdxs \, \lg_1 \td >0, \Tr \ts \Phdxs \, \lg_6 \td >0
\end{array}
\eeq
The meaning of these three conditions becomes clearer when one writes explicitly 
$\Phdxs$
\beq\label{matstep2}
\Phdxs =
\ts \begin{array}{lll}
\phd_3\,'  +  \frac{1}{\sqrt{3}} \phd_8\,'  &
\ts \phd_1\,'  -i \phd_2\,'  \td \esp^{3i\ag } &  
\ts \phd_4\,'  -i \phd_5\,'  \td \esp^{3i(\ag  + \bg )}\\
\ts \phd_1\,'  +i \phd_2\,'  \td \esp^{-3i\ag }& 
-\phd_3\,'  + \frac{1}{\sqrt{3}}  \phd_8\,'  &
\ts \phd_6\,'  -i \phd_7\,'  \td \esp^{3i \bg }\\
\ts \phd_4\,'  +i \phd_5\,'  \td  \esp^{-3i(\ag  + \bg )}& 
\ts \phd_6\,'  +i \phd_7\,'  \td\esp^{-3i\bg }&  
\qquad - \frac{2}{\sqrt{3}} \phd_8\,'
\end{array} \td
\eeq
For example, $\ag (x)$ and $\bg (x)$ can be fixed requiring that
\beq\label{alpha}
\esp^{3i\ag (x)} =\frac{\phd_1\,' (x) +i \phd_2\,' (x)}
{\sqrt{\phd_1\,'^2 (x) + \phd_2\,'^2 (x)}}
\eeq
\beq\label{beta}
\esp^{3i\bg (x)} =\frac{\phd_6\,' (x) +i \phd_7\,' (x)}
{\sqrt{\phd_6\,'^2 (x) + \phd_7\,'^2 (x)}}
\eeq
making the elements $(1,2)$ and $(2,3)$ of $\Phdxs$ real, i.e. satisfying
conditions 1) and 2).
The condition 3) in (\ref{step2req}) is introduced to eliminate the sign
ambiguities $ \pm \ts \phd_1\,' (x) +i \phd_2\,' (x) \td$ and
$ \pm \ts \phd_6\,' (x) +i \phd_7\,' (x) \td$. 
In conclusion, after this second step, we have obtained the
transformation $W(x)=V(x) \Og (x)$ which fixes uniquely the gauge up to
the center symmetry $Z_3$.
In the $SU(N)$ case, the symmetry must be reduced from $U(1)^{N-1}$ to $Z_N$
by fixing $(N-1)$ phases. This can be accomplished, in a simple and elegant way, 
by requiring that the $(N-1)$ sub--diagonal elements of $\Phdxs$
be real positive. Moreover, with this choice of constraints, the conditions 
defining the monopoles turn out to be particularly simple.

Two last important remarks concern the accidental degeneracy of
$\mu_1$ or $\mu_2$ in (\ref{eigeq}) and the sign arbitrariness in the
eigenvectors $\phi^{(j)}$. In the described procedure we have assumed
that the two lowest eigenvalues $\mu_1$ and $\mu_2$ of the Laplacian
operator $\Dg$ are both non-degenerate. If either one is degenerate, 
the gauge fixing can not be carried out in an unambiguous way; however 
these cases are really exceptional and in the numerical simulations
never occur. Moreover this degeneracy is easy to
check and so these events can be detected. The second point is about 
the arbitrariness in the eigenvectors $\phi^{(j)}$, since they are defined 
up to a global scale factor and a global sign. The two--step method to fix the 
gauge makes use of the orientation of the vectors $\phux$ and $\phdx$
in color space. So, while the rescaling can not give rise to any
ambiguity in the procedure, the freedom in the choice  
of the global sign can. This freedom can be eliminated with a conventional
choice on $\phi^{(j)}$.

\section{Local gauge ambiguities}

Local defects may occur in the Laplacian Center Gauge fixing
procedure. In this section we discuss how they show up and how
they can be associated with monopoles and center vortices. This association
is consistent with the initial proposal by 't Hooft of identifying
gauge fixing defects and topological features of the gauge fixed
theory. Moreover, in this approach, we will make apparent the close
relation which exists between monopoles and center vortices in the
Laplacian Center Gauge. Let us start by showing the conditions that give rise 
to defects at each step of the gauge fixing procedure.

{\bf{Step 2 ill--defined}}\\
In {\bf{Step 2}} we fix the $U(1)^{N-1}$ symmetry by looking for
gauge transformations $V(x)$ in the Cartan subgroup that make some color
components of $\phdxs$ vanish. This is equivalent to requiring that
particular entries in the complex matrix $\Phdxs$ be real positive. If it happens that, 
at some point $x$, any of these entries is zero, the second step can not be 
carried out completely. Let us illustrate the argument by considering $SU(3)$.
The conditions (\ref{step2req}) specify $V(x)$ via the phases $\ag (x)$ and
$\bg (x)$ given by (\ref{alpha}) and (\ref{beta}). If, at some point
$x$, one of the following cases takes place
\beq\label{step2ill}
\begin{array}{l}
\mbox{1) } \phd_1\,' (x) -i \phd_2\,' (x) =0\\
\mbox{2) } \phd_6\,' (x) -i \phd_7\,' (x) =0
\end{array}
\eeq
the corresponding phase -- $\ag(x)$ for 1) and $\bg(x)$ for 2) -- 
is not defined. Thus, at point $x$ the gauge symmetry cannot be
reduced to $Z_3$ and one of the two $U(1)$ subgroups is left unfixed.
The conditions 1) or 2), and more generally the corresponding ones for $SU(N)$, 
set two constraints, and so the points $x$ where the remaining gauge freedom
is enlarged from $Z_N$ to include a continuous $U(1)$ group form 2--dimensional 
surfaces in the 4--dimensional space--time.
For $SU(2)$, these surfaces are defined by the collinearity of $\phi^{(1)}$
and $\phi^{(2)}$.

{\bf{Step 1 ill--defined}}\\
In {\bf{Step 1}} we reduce the gauge symmetry from $SU(N)$ to
$U(1)^{N-1}$ by diagonalizing $\Phux$ with a
conventional ordering for the eigenvalues along the diagonal.
If, at some point $x$, two eigenvalues happen to be equal,
the conventional ordering is no longer unique and the procedure becomes
ill-defined at that point. For instance, let us consider
$SU(3)$. The traceless matrix $\Phuxp$ can be written in the form  
\beq\label{su3ex}
\Phuxp =
\frac{1}{\sqrt{3}}
\ts \begin{array}{lll}
2 a(x) + b(x) &              &                \\
              & -a(x) + b(x) &                \\
              &              & -a(x) - 2 b(x) 
\end{array} \td
=  a(x) \lg_3 '+ b(x) \lg_8
\eeq
where $a(x)$ and $b(x)$ are non-negative and 
$\lg_3 '=\frac{1}{2} (\sqrt{3} \lg_3 + \lg_8)$ ($\lg_3$ and
$\lg_8$ are the diagonal Gell--Mann matrices). In writing
(\ref{su3ex}) we have assumed a conventional decreasing ordering along
the diagonal for the eigenvalues.
$\Phuxp$ has two equal eigenvalues in one of the following two cases:
\beq\label{step1ill}
\begin{array}{l}
\mbox{1) } a(x) = 0\\
\mbox{2) } b(x) = 0
\end{array}
\eeq
When one of these occurs, the residual gauge symmetry is $SU(2)\times
U(1)$ (or more precisely $U(2)$) instead of $U(1)^2$. Suppose that case 1) takes place, then
$\Phuxp$ is invariant under gauge transformations generated 
by $\lg_8$, $\lg_1$, $\lg_2$ and $\lg_3$ and the last three matrices
generate an $SU(2)$ subgroup. For the generic $SU(N)$ group, the
symmetry is $SU(2)\times U(1)^{N-2}$ \footnote{More precisely, 
by this notation we mean that
the group symmetry is generated by the elements of the Cartan subalgebra plus
two more $su(N)$ matrices which, together with one of the already
considered diagonal generators, form an $su(2)$ subalgebra.}.
Let us now show that these points $x$ where the first
step can be performed only partially and the gauge symmetry is reduced 
to $SU(2)\times U(1)^{N-2}$, form 1--dimensional strings in 
4--dimensional space--time. Imagine that the matrix $\Og (x)$
which diagonalizes $\Phux$ is computed in two stages. In the first stage, we partially 
diagonalize $\Phux$ so as to leave only two (symmetric) off--diagonal elements different
from zero. In the second stage, an $SU(2)$ transformation is
considered which makes these last two off--diagonal elements vanish.
This second stage is equivalent to the diagonalization of an $su(2)$ matrix 
$s(x)=\sum_1^3 s_k(x) \sg_k$. Without loss of generality, 
we can always suppose that the $(N-2)$ diagonal elements found after the first
stage and not touched by the $SU(2)$ transformation of the second
stage, are all different, and that the equality of two
eigenvalues of $\Phux$ can show up only in the second stage (this means 
that, in general, the position of the non-zero off--diagonal elements in
the partially diagonalized matrix depends on $x$ and on the
conventional ordering). Then the equality of two eigenvalues takes 
place at the points $x$ where $s_1(x)=s_2(x)=s_3(x)=0$: 
since three constraints must be satisfied, the points $x$ form
1--dimensional strings in 4--dimensional space--time.

Having established that the local gauge ambiguities at Step 1 and 2
have co-dimension 3 and 2 respectively, we now proceed to associate them
with monopoles and center vortices.
Let us first show that the gauge fixing defects of 
{\bf{Step 2 ill--defined}} can be associated with center vortices. Again 
we consider $SU(3)$ and the extension to $SU(N)$ is straightforward.

Suppose that, at some point $x_0$, case 1) of (\ref{step2ill}) takes
place. Let us define 
$\phi_{1,2}^{(2)} \, ' (x) = \phd_1\,' (x) +i \phd_2\,' (x)$ and
consider its Taylor expansion around $x_0$:
\beq
\phi_{1,2}^{(2)} \, ' (x) = (x-x_0 ) \cdot \nabla 
\phi_{1,2}^{(2)} \, ' (x_0 ) + {\cal{O}} (x-x_0 )^2
\eeq
obtained keeping in mind that $\phi_{1,2}^{(2)} \, ' (x_0)=0$. 
Consider then a plane passing through $x_0$. Moving on 
this plane in the neighborhood of $x_0$, $\phi_{1,2}^{(2)} \, ' (x)$ has, up
to some complex rescaling factor, a hedgehog shape that we can parametrize in 
polar coordinates
\beq
\phi_{1,2}^{(2)} \, ' (r, \theta ) \simeq f(r,\theta)\, \esp ^{i\theta}
\eeq
with $f(r,\theta)$ a real positive function. If we now impose the gauge
fixing conditions (\ref{step2req}) and we take into account 
(\ref{matstep2}), we obtain that, in the neighborhood of $x_0$ and
on the considered plane, the {\bf{Step 2}} gauge transformation is
given by $V(x)={\mbox{diag}}(\esp^{-i\frac{2\theta}{3}},
\esp^{i\frac{\theta}{3}},\esp^{i\frac{\theta}{3}})$. In this
argument we do not consider the $U(1)$ subgroup parametrized by 
$\bg (x)$, since its contribution can be factored out. Applying the
gauge transformation $V$
\beq
A_{\mu} \longrightarrow V( A_{\mu}+\frac{i}{g} \partial _{\mu})
V^\dagger
\eeq
we obtain that, moving in the plane in the neighborhood of $x_0$,
the gauge field gains the term
\beq
\frac{i}{g} V \partial _{\mu} V^\dagger =
-\frac{1}{3gr}\, \lg_3 '\, \vec{\esp}_{\theta}
\eeq
If we now integrate over a closed path $C$ around $x_0$ we have
\beq
\oint_C \frac{1}{3gr} \,r \, dr\, d\theta =\frac{2\pi}{3}
\eeq
This means that a Wilson loop belonging to the considered plane and
encircling $x_0$ gets a non-trivial factor 
$\esp^{i\frac{2\pi}{3}} 1$~\hskip-3.65mm~I with respect to the 
center group $Z_3$. A similar discussion can be carried out for case
2) of (\ref{step2ill}). According to the presented argument, we
establish a connection between center vortices and gauge fixing
defects arising from  {\bf{Step 2}}.\\
As for those arising from {\bf{Step 1}},
't Hooft already has shown how such gauge fixing defects 
can be identified with monopole world--lines\cite{thooft1}. 
He has also explained that, in $SU(N)$, all monopoles are composite of $(N-1)$
elementary ones corresponding to the degeneracy of two neighbouring eigenvalues.
Here we discuss the feature that, in the Laplacian Center Gauge, center
vortices and monopoles turn out to be closely related in a unified
description. Take the $SU(3)$ theory and consider the 2--dimensional surface $\Sigma$
of the center vortices defined by $\phd_1\,' (x)=0$, $\phd_2\,' (x)=0$.
Similarly to equation (\ref{su3ex}), we can write the diagonal
elements of $\Phdxs$ in the form $c(x) \lg _3 '+d(x) \lg_8$. Now     
$c(x)$ and $d(x)$ can be also negative, since the permutation freedom has 
already been used to order the eigenvalues of $\Phu$. 
At every point $x$, $a(x)$ and 
$c(x)$ may have identical or opposite signs. Let us restrict ourselves to 
$x\in \Sigma$, and suppose that $a(x)$ and $c(x)$ have identical sign in
some parts of $\Sigma$, opposite sign in others. By
continuity, this implies that these patches must be separated by
1--dimensional closed strings where $a(x)=0$ or $c(x)=0$, 
the former of which defines a monopole world--line. Thus the
2--dimensional surface of center vortices contains embedded monopole
world--lines. Moreover, all monopole world--lines are embedded in
center vortex surfaces. 
To see this, consider an $SU(2)$ static monopole at some point $x_0$ in
3--dimensional space. By definition, $\phu$ vanishes at the monopole core $x_0$,
and in its neighborhood it has a hedgehog shape in color space. So there will 
necessarily exist some
direction along which $\phux$ is parallel to $\phdx$ -- which is the
condition identifying center vortices in $SU(2)$ --. Thus
the core of a monopole is nested in the 1--dimensional closed string
of a center vortex, which appears as two half-Dirac strings. 
Considering now the time evolution also, we see
that a monopole world--line is always embedded in a 2--dimensional
surface of center vortices. Since, in general, 
the pattern of the gauge symmetry promotion at the monopole core is  
$U(1)^{N-1} \longrightarrow SU(2)\times U(1)^{N-2}$ -- i.e. one
of the $U(1)$ subgroups is promoted to $SU(2)$ --, the argument
given here for $SU(2)$ can be extended to a generic $SU(N)$.
Fig.1 sketches the embedding of monopole world--lines
in center vortex surfaces just discussed.
\begin{figure}[h]
\begin{center}
\psfig{figure=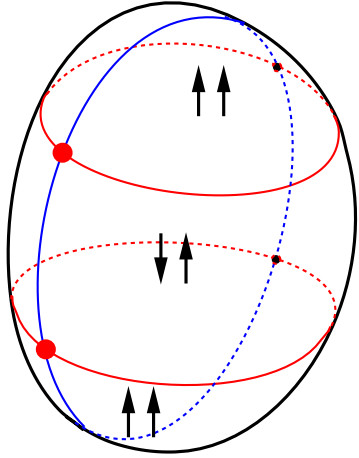,height=6cm}
\vspace{0.3cm}
\caption{The connection between center vortices and monopoles:
the (horizontal) monopole world--lines separate two patches of center vortex 
surface with opposite eigenvector orientations; each monopole is attached to two
center vortex strings.}
\end{center}
\label{sketch}
\end{figure}

This close connection between monopoles and center vortices has actually
been observed on the lattice \cite{green+1,green+2}, and even presented as a puzzle \cite{Stack_Osaka}: 
on a cooled $SU(2)$ lattice configuration, almost all Abelian monopoles, 
identified as a $3d$ cube through which one measures a $2 \pi$ magnetic flux, 
are attached to two opposite center vortex plaquettes. 
This is precisely the situation we describe: 
as correctly anticipated in \cite{green+1,green+2},
Abelian monopoles and anti-monopoles are like alternating beads on a necklace
formed by the center vortex string.

\section{Numerical results}

We have performed numerical simulations to investigate the role of the 
center degrees of freedom in the $SU(2)$ and $SU(3)$ lattice gauge
theories. According to the discussion of the two previous sections, we 
have made use of the Laplacian Center Gauge fixing to reduce the
symmetry to the center subgroup. In this approach -- considering for
instance $SU(3)$ -- we have described how center vortices can be
detected by looking for the points $x$ where, for example, 
$\phi_{1,2}^{(2)} \, ' (x) =0$ or, equivalently, $\phdx$ describes a
$2\pi$ rotation in color space around the $\lg_3 '$ axis, moving along 
a closed contour encircling $x$. The method is the same for a generic
group $SU(N)$. As a by-product, monopole world--lines are identified by the
points $x$ where two eigenvalues of $\Phuxp$ are equal; but our aim here is 
to study the center degrees of freedom and not the Abelian ones.
Numerical simulations are performed on a lattice and so the
eigenvectors of the adjoint Laplacian operator can be computed only at 
discrete points. This implies that an interpolation procedure must
be defined to detect center vortices and, if desired, monopoles. We
have tried several interpolation methods  in order to investigate the
robustness of the results with respect to different
choices. Unfortunately, the arbitrariness introduced by this necessary 
step seems large at the lattice spacings we considered. 
We have found only one plaquette-based interpolation scheme which guarantees
that center vortex surfaces are closed. It is identical to the construction
of \cite{Rajantie}, eq.(10).
Remarkably, this interpolation scheme gives center vortices at locations
which {\em coincide} with the $P$-vortices obtained after center projection by 
Greensite and collaborators \cite{green1}. Following this prescription,
every gauge fixed link $U_\mu (x)$ is decomposed as the product of two 
parts
\beq\label{decomp}
U_\mu (x) = Z_\mu (x) \cdot U_\mu ' (x)
\eeq
where $Z_\mu (x)$ is the center projected link living in $Z_N$ and 
$U_\mu ' (x)$ is the coset link taking values in $SU(N)/Z_N$. For
instance, this splitting can be carried out by requiring that
$|\mbox{arg(Tr}(U_\mu ' (x)))| \leq \pi/N$. Thus, starting
from an $SU(N)$ gauge fixed configuration, one builds up two projected
configurations, made of the $Z_\mu (x)$ and the $U_\mu ' (x)$ links. Then, if
$W(C)$ is a Wilson loop along the closed contour $C$, making use of 
the decomposition (\ref{decomp}), one can write
\beq\label{WC}
W(C)=\sigma (C) \, W' (C) = 
\qs \prod_{p\in \Sigma}\, \sigma (p) \qd W' (C) 
\label{WLdecomp}
\eeq
where $\sigma (p)$ is the plaquette built from center links $Z_\mu(x)$, and
$W' (C)$ and $\sigma (C) \equiv \prod_{p\in \Sigma}\, \sigma (p)$ are
the Wilson loops evaluated with the coset links 
$U_{\mu}' (x)$ and with the center links $Z_{\mu} (x)$ respectively. 
$\prod_{p\in \Sigma}$ is the product over all the plaquettes $p$
belonging to a surface $\Sigma$ supported on $C$; the value of 
$\sigma (C)$ does not depend on the choice of $\Sigma$ and, for
simplicity, we can suppose that it is the planar surface bounded by 
$C$. 
Since we have fixed a gauge where $U_{\mu}'$ is smooth, we expect
that, to a reasonable approximation, $W(C)$ has a non-trivial value with
respect to $Z_N$ if and only if $\sigma (C)$ does. In this approach, a non-trivial
value for $\sigma (p)$ is the ``signal'' for a center vortex.
We stress again that the vortices identified in this way {\em coincide}
with interpolated gauge defects, which brings center projection on firmer
theoretical ground.

We have collected 1000 $SU(2)$ configurations at three different
values of the coupling constant $\bg=2.3$, $2.4$ and $2.5$ on a $16^4$
lattice; for $SU(3)$ we have generated 500 configurations on a $16^4$
lattice at $\beta=6.0$.  

The following figure shows our measurement of the Creutz ratios 
$\chi(R)=-\ln (\langle W(R,R)\rangle \langle W(R-1,R-1)\rangle 
/\langle W(R,R-1)\rangle^2)$ for $SU(2)$ at $\beta=2.4$ ($W(R,T)$ is
the Wilson loop $R\times T$). 
\begin{figure}[h]
\begin{center}
\psfig{figure=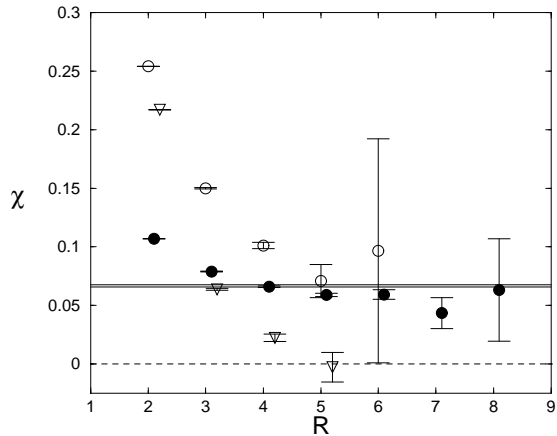,height=6cm}
\caption{Creutz ratios at $\bg=2.4$. Empty circles refer to $SU(2)$,
  full circles to $Z_2$ and triangles to the coset $SU(2)/Z_2$.
The continuous strip is the value of the string tension in the literature \protect\cite{teper}.}
\end{center}
\end{figure}
Empty circles refer to $SU(2)$, full circles to center projection after Laplacian
Center Gauge fixing, and triangles to the coset part. The continuous strip is the
value in the literature \cite{teper,bali1} of  
the $SU(2)$ string tension for the chosen set of
parameters. These numerical results show, on one hand, the flattening of 
the Creutz ratios in the $Z_2$ sector and, on the other hand, the
vanishing of the Creutz ratios computed with the coset links. We have
obtained similar behaviour for the other two values of $\beta$: $2.3$ 
and $2.5$. We have also carried out the same study for the $SU(3)$ lattice 
gauge theory. The next figure shows our results for the Creutz
ratios in the $Z_3$ sector after Laplacian Center Gauge fixing of 500
configurations. 
\begin{figure}[h]
\begin{center}
\psfig{figure=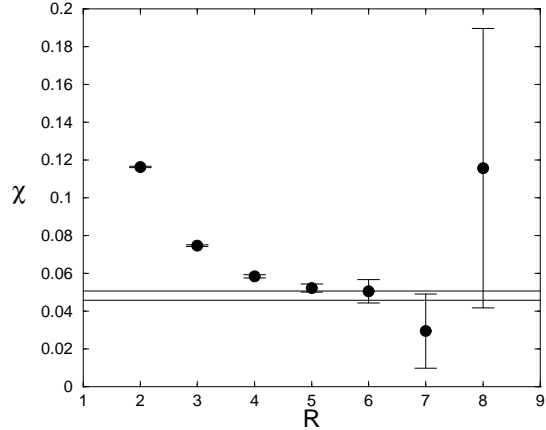,height=6cm}
\caption{Creutz ratios at $\bg=6.0$ in the center sector $Z_3$ of
  $SU(3)$.
The continuous strip is the value of the string tension in the literature \protect\cite{bali2}.}
\end{center}
\end{figure}
The continuous strip is the value in the literature\cite{bali2} of the 
$SU(3)$ string tension at the considered set of parameters. Also in
this case, one clearly sees a nice flattening to a non-vanishing
value of the Creutz ratios evaluated with the center projected
links. 

Therefore, our $SU(2)$ and $SU(3)$ results are qualitatively similar to
those previously obtained in Direct Maximal Center gauge in $SU(2)$.
They confirm the finding that the center-projected ensemble of gauge defects confines,
with a string tension close to that of the original non-Abelian theory,
whereas the coset ensemble does not. One difference however is apparent:
the Creutz ratios tend to an asymptotic value much more slowly in
Laplacian gauge than in DMC gauge. This is caused by the presence of
many more close pairs of center vortices. Indeed, the vortex density
is much higher, by a factor 3 to 5, than that measured in DMC gauge.
A similar increase in the density of Abelian monopoles was observed
previously in the Laplacian Abelian gauge \cite{vds}.
One may consider this a practical nuisance, since these additional
vortex pairs make the extraction of the projected string tension more
noisy. One may instead consider that the center-projected ensemble
represents the original one more closely, since it partially reproduces
the short-distance increase of the force.

Finally, one can compare DMC and Laplacian Center gauges from the point of
view of the computer effort. Because two eigenvectors of the adjoint
Laplacian must be computed, one may get the impression
that the Laplacian gauge is computationally expensive. That impression
is misleading. Using the public-domain package ARPACK \cite{ARPACK}
to solve the eigenvalue problem, the computer time needed to gauge
fix one $16^4$ configuration is about the same as for 50 Monte Carlo 
sweeps in the case of $SU(2)$, and 300 to 500 for $SU(3)$. 
This is far less than required to fix to DMC gauge iteratively.

\section{Interpretation of the numerical results}

The figures presented in the previous section show  good agreement
between the Creutz ratios evaluated from the center projected ensemble of gauge defects
and the value of the string tension in the literature for $SU(2)$ and
$SU(3)$ at the same $\bg$. However, the importance of this numerical agreement
should not be overestimated. Numerical simulations are performed at a
finite value of the lattice spacing, and we see no reason to believe
that, if lattice artifacts are negligible for the $SU(N)$ gauge theory, their
effect is equally small on the results obtained from the center
projected model. Nevertheless, our conjecture is that, even if
the value of the string tension in the center sector is appreciably
modified by lattice artifacts at finite lattice spacing, this dependence has to
vanish in the continuum limit. The observation of such a behaviour
would be a robust confirmation of the relevance of the center degrees
of freedom in the confinement mechanism. In order to investigate this
issue, we have considered three different lattice Laplacian operators
to fix the gauge. They differ by terms which vanish in the continuum
limit, i.e. higher derivatives or irrelevant operators. 
In practice, these new Laplacians have been obtained very simply, by smearing
the links $U_\mu$ and substituting smeared links in the construction of 
the Laplacian (\ref{adjlap}). Specifically, we have 
considered $0$, $1$ and $5$ smearing steps (staple weight = $0.5$) on the
$U_\mu$ links. We stress that the smeared links are used only to obtain different
gauge fixing operators; the gauge transformations are always applied
to the non-smeared link configurations, 
as well as the center projection and measurements. 

The same sample of configurations has been fixed 
in each one of the three gauges, and the string tension has been
measured in the center sector after center projection each time. 
In order to estimate more accurately the string tension from the $Z_\mu$ links, 
we have constructed smeared Wilson loops, where the spatial sides
are made of links recursively smeared with a fixed time coordinate.
This procedure is identical to that commonly used in the measurement
of the $SU(N)$ static potential. It reduces the contribution of
excited states, thus improving the statistical accuracy on the string tension.
Although in the center projected theory, the existence of a transfer matrix
is doubtful, we verified that spatial smearing did not introduce a measurable
bias.
From the Wilson loop data, we have extracted the $Z_N$ string tension 
using the same fitting procedure as employed for the $SU(N)$ case,
with an ansatz of the form $V_0+\sg R -\frac{e}{R}$ for the static potential $V(R)$
\footnote{We gratefully acknowledge G. Bali for providing us with a data analysis
program similar to that used in \cite{bali1,bali2}, so that possible systematic 
errors in the evaluation of the $Z_N$ and the $SU(N)$ string tensions largely
cancel out when considering their ratio.}

To illustrate our results, we show in the following figures the
$Z_2$ and $Z_3$ Creutz ratios measured in the three gauges. The gauge dependence
of the projected string tension is dramatic.
\begin{figure}[h]
\begin{center}
\psfig{figure=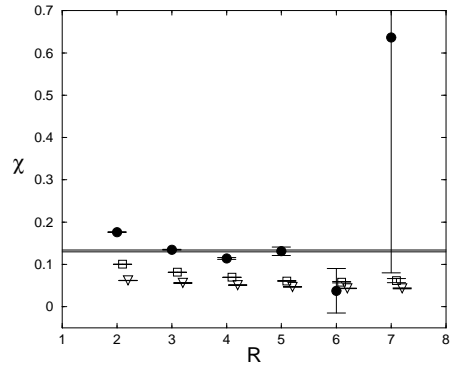,height=5.cm}
\caption{$Z_2$ Creutz ratios at $\bg=2.3$ in the various gauges. Full
  circles refer to the usual Laplacian, squares to the once-smeared and
  triangles to the 5-times smeared Laplacian. The continuous strip is the
  value of the string tension in the literature \protect\cite{teper}.}
\end{center}
\end{figure}
\begin{figure}[h]
\begin{center}
\psfig{figure=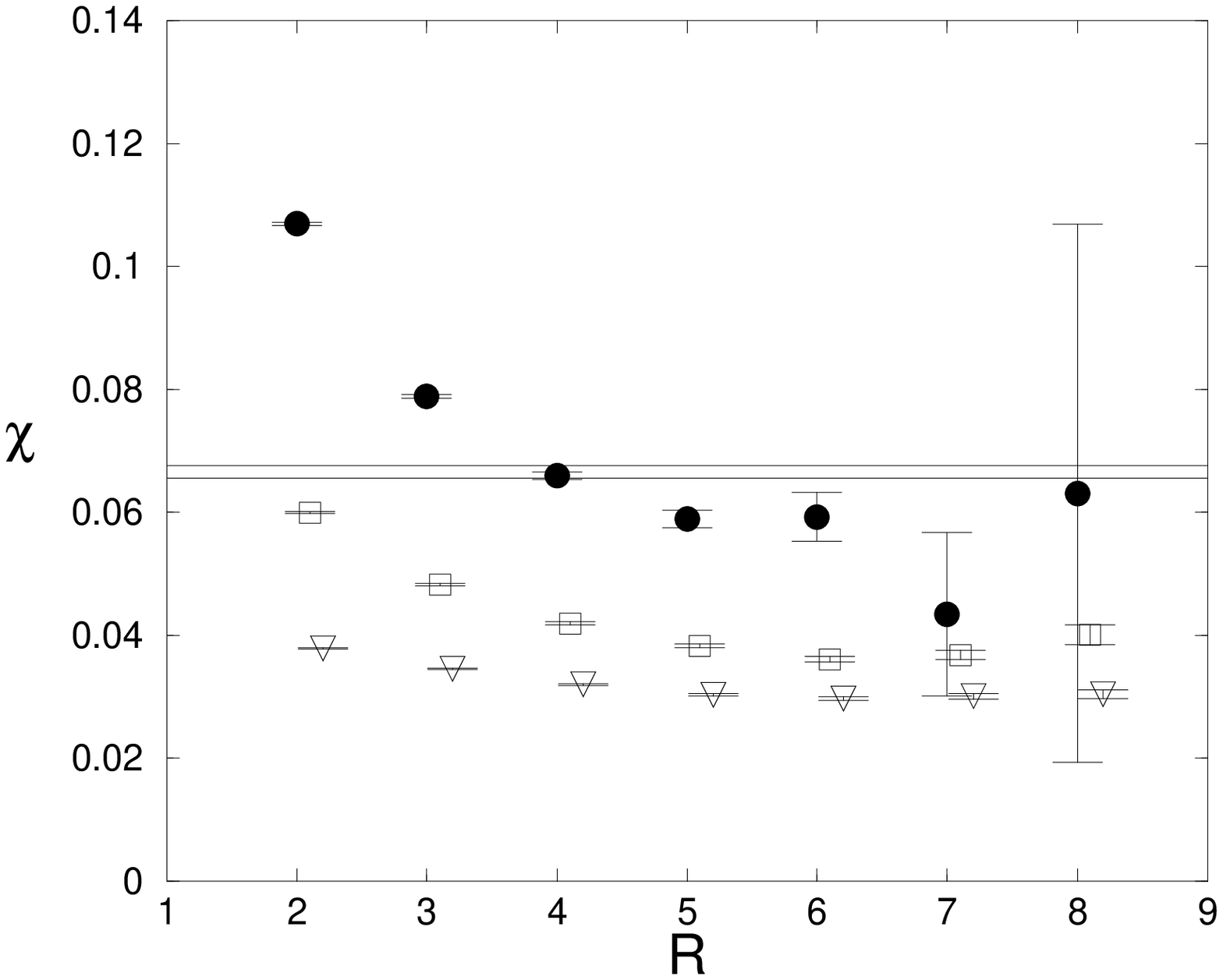,height=5.cm}
\caption{$Z_2$ Creutz ratios at $\bg=2.4$ in the various gauges. Full
  circles refer to the usual Laplacian, squares to the once-smeared and
  triangles to the 5-times smeared Laplacian. The continuous strip is the
  value of the string tension in the literature \protect\cite{teper,bali1}.}
\end{center}
\end{figure}
\begin{figure}[h]
\begin{center}
\psfig{figure=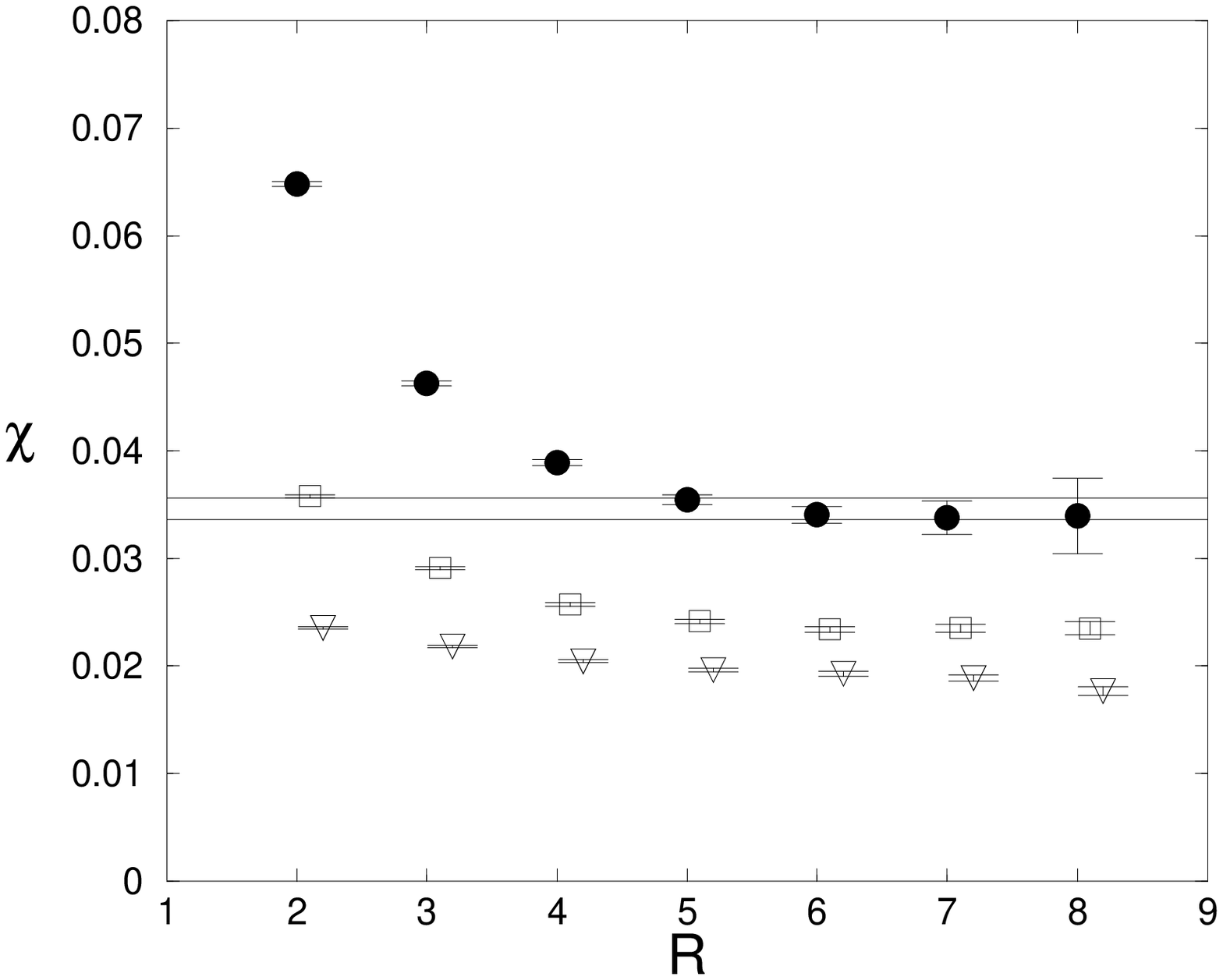,height=5.cm}
\caption{$Z_2$ Creutz ratios at $\bg=2.5$ in the various gauges. Full
  circles refer to the usual Laplacian, squares to the once-smeared and
  triangles to the 5-times smeared Laplacian. The continuous strip is the
  value of the string tension in the literature \protect\cite{teper,bali1}.}
\end{center}
\end{figure}
\begin{figure}[h]
\begin{center}
\psfig{figure=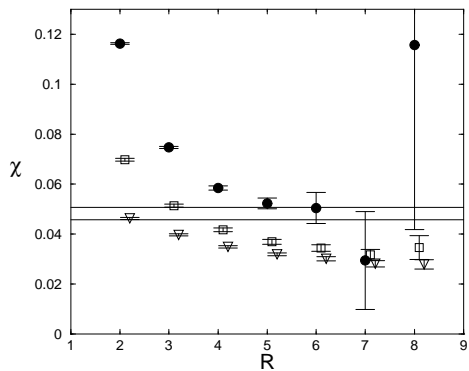,height=5.cm}
\caption{$Z_3$ Creutz ratios at $\bg=6.0$ in the various gauges. Full
  circles refer to the usual Laplacian, squares to the once-smeared and
  triangles to the 5-times smeared Laplacian. The continuous strip is the
  value of the string tension in the literature \protect\cite{bali2}.}
\end{center}
\end{figure}

The string tensions $\sg_i$ ($i=0,1,5$ is the number of smearing steps
performed on the links $U_\mu$ entering in the Laplacian operator
eq.(\ref{adjlap})) obtained
from our 1000 $SU(2)$ configurations at $\bg=2.3$, $2.4$, $2.5$ in the
three different Laplacian Center Gauges are then compared with the
$SU(2)$ string tension $\sg_{SU(2)}$ reported in the literature. 
We denote by $R_i$ the ratio $R_i =\sqrt{\sg_i/\sg_{SU(2)}}$.
The following table summarizes our results:

\begin{center}
\begin{tabular}{||c||c|c|c||}
\hline
\qquad {} \qquad & 
$\qquad \bg =2.3\qquad$ & $\qquad\bg =2.4\qquad$ & $\qquad\bg =2.5\qquad$ \\
\hline
$R_0$ & $0.813(23)$ & $0.860(20)$ & $0.978(18)$\\
$R_1$ & $0.592(12)$ & $0.720(11)$ & $0.804(12)$\\
$R_5$ & $0.547(8) $ & $0.653(7) $ & $0.739(11)$\\
\hline
\end{tabular}
\end{center}

As the number of smearing steps $i$ of the links entering the Laplacian
increases, the projected string tension $\sg_i$ quickly decreases.
This effect has nothing to do with the variation of the string tension
(or absence thereof) under cooling or smearing. Here the gauge configuration
is always the same unsmeared one. Only the operator used for gauge fixing
changes. In this case, the reduction of the projected string tension is
qualitatively easy to understand. The Laplacian made of
smeared links becomes increasingly blind to short-range features of the gauge
field, so that,
after gauge fixing, the factorization of the Wilson loop eq.(\ref{WLdecomp})
into a smooth non-Abelian part $W'(C)$ and a disordered center part $\sigma(C)$
is less effective. The smooth part $W'(C)$ is less smooth and carries some disorder,
while the center part $\sigma(C)$ is less disordered, showing a reduced string
tension. However, as $\beta$ increases, the physical smearing radius $\rho$ 
shrinks to zero, and one expects full identification of all center vortices.
Our Table shows that the numbers in each row increase with $\beta$ towards 1.
The more reliable the identification of the gauge defects in 
the Laplacian Center Gauge, the closer the matching between the projected
and the full string tensions. Thus, our numerical evidence supports the idea
that a complete matching takes place in the continuum limit, i.e. center dominance
(see Note Added). 

If, on the other hand, the amount of smearing in our covariant Laplacian was adjusted as a function of the lattice
spacing so that the physical smearing radius $\rho$ would remain constant, then
one would achieve a continuum limit where, presumably, the center-projected
string tension would be a fraction of the non-Abelian one. This fraction
should decrease as the smearing radius $\rho$ grows. When $\rho$ reaches 
the physical size of a center vortex, the smeared Laplacian becomes blind
to center vortices and the decomposition (\ref{WLdecomp}) becomes completely
ineffective. This simple reasoning shows that the covariant operator used for
the gauge fixing must be {\em local} for the center dominance scenario to be correct.
Indeed, according to our Table, the more local the operator, the more faithful
center projection appears to be at any finite lattice spacing.

A similar analysis has been carried out for $SU(3)$ and the following
table shows the corresponding results:

\begin{center}
\begin{tabular}{||c||c||}
\hline
\qquad {} \qquad & 
$\qquad \bg =6.0\qquad$\\
\hline
$R_0$ & $0.93(6)   $\\
$R_1$ & $0.818(32) $\\
$R_5$ & $0.739(24) $\\
\hline
\end{tabular}
\end{center}

The trend is similar to $SU(2)$. Furthermore, the ratios $R_i$ are very close
to those measured in $SU(2)$ at $\beta=2.5$. Following the argument
presented above, this indicates that the typical vortex size, in
lattice units, is the same in both cases. With 
$a(SU(3),\beta=6.0) \approx 0.1$ fm and $a(SU(2),\beta=2.5) \approx 0.085$ fm,
we obtain that the ratios of $SU(3)$ over $SU(2)$ center vortex sizes is about
$0.1 / 0.085 \sim 1.18$. Such a slight increase is consistent with the
expected increase of the adjoint string-breaking distance \cite{Owe},
which is connected to the center vortex size \cite{CVsize}.

\section{Summary and discussion}

The standard iterative local maximization methods used to fix the gauge on
the lattice are ambiguous: they stop when {\em any} local maximum has been
reached. Because of this ambiguity, one tends to mistrust measurements 
performed in such ill-defined gauges, as well as the physical models based
on these measurements. This applies in particular to the scenario of vortex
dominance, according to which the string tension of the non-Abelian theory
can be exactly reproduced by considering only its center degrees of freedom,
identified after gauge-fixing. Evidence for this scenario comes exclusively
from lattice studies using ambiguous gauges. Some amount of counter-evidence
\cite{K&T,borny} has also been reported, again in similar ambiguous gauges.

Motivated by our skepticism, we have constructed, by generalizing the approach
of \cite{vink}, a gauge which smooths the adjoint $SU(N)/Z_N$ field like the 
usual Maximal Center Gauge (MCG), but has no ambiguity. After gauge fixing to
this Laplacian Center Gauge, a remaining local $Z_N$ gauge freedom subsists.
The local gauge defects which appear in this gauge are of two types:
co-dimension 2, where the remaining gauge freedom is enlarged from $Z_N$ to 
$U(1)$, i.e. to accommodate a $U(1)$ subgroup; and co-dimension 3, where it is 
further enlarged to accommodate an $SU(2)$ subgroup. 
These two types of defects can be identified with center vortices and
Abelian monopoles respectively, with the latter embedded in the former.
Thus we provide a natural, unified description of these two objects which
had been considered as alternative choices of effective degrees of freedom.

We have numerically implemented the Laplacian Center Gauge for $SU(2)$ and
$SU(3)$. In addition to being free of the ambiguities which plague the
Maximal Center Gauge, our gauge is also computationally cheaper.
The gauge defects obtained by local interpolation of the Laplacian eigenvectors
coincide with the $P$-vortices obtained by center projection.
A measurement of Creutz ratios in the center-projected $Z_2$ and $Z_3$ 
ensembles superficially confirms earlier observations made in MCG, albeit
with a higher density of center vortices: the center-projected ensemble
confines like the original one, while the coset $SU(N)/Z_N$ ensemble does not.
Upon closer scrutiny however, the center-projected string tension is
smaller than the original one. The difference between the two can be 
varied by arbitrary amounts, by adding higher derivative terms to the
Laplacian used for the gauge fixing. Nevertheless, we have shown 
evidence that this difference decreases as the continuum limit 
of the lattice theory is approached. The gauge dependence of the center-projected
string tension, clearly visible at finite lattice spacing $a$, appears to go away 
as $a \rightarrow 0$. Therefore, while our study calls attention to lattice 
artifacts in the center projection, it gives support to the center dominance scenario.

Since the center degrees of freedom (d.o.f.) reproduce the non-Abelian 
confinement, one may wonder if the center-projected model can also reproduce
other non-perturbative features of the non-Abelian one, like chiral symmetry
breaking or topological susceptibility. Indeed, one would expect this to be
the case, since the coset $SU(N)/Z_N$ ensemble appears deprived of all these
non-perturbative features \cite{phmax1}. However, one encounters two difficulties
when trying to measure the appropriate observables in the $Z_N$ model:
lattice artifacts and non-positivity of the transfer matrix.

We have already emphasized how lattice artifacts, in the choice of the
discretized Laplacian or in the interpolation/center projection step, can affect the 
matching of the $Z_N$ string tension with the $SU(N)$ one. Lattice artifacts
modify the center vortex density so strongly that we cannot really quote
a value for this important quantity. They also strongly influence the
measurement of the chiral condensate in the $Z_N$ ensemble\cite{phmax2}.
Naturally, they also make it conceptually difficult to define a topological
charge on the lattice, especially for a discrete $Z_N$ theory.

Another, fundamental, difficulty comes from the non-locality of the gauge
condition. After gauge fixing, all gauge links are correlated with each other,
and this correlation persists after center projection. Furthermore, it may
not die out exponentially at large distances in the center-projected ensemble,
making it impossible to define a local effective Hamiltonian and a positive
transfer matrix. Indeed, we have observed symptoms of this disease when 
attempting to measure glueball masses in the $Z_N$ ensemble: in several cases,
correlations at large distances would become significantly negative,
and impossible to interpret as the Euclidean-time propagator of a superposition
of eigenstates of a Hamiltonian. Difficulties have also been reported for
the Abelian projection \cite{Miyamura,Stack}.
Symptoms may even be present for the string tension, since in Ref.\cite{vds},
Creutz ratios measured after Abelian projection in the Maximal Abelian Gauge
appear to {\em increase} with distance; a proper, normal decrease was 
restored when using instead the Laplacian Abelian Gauge. 
We also observed a similar increasing behaviour for the $Z_N$ Creutz ratios 
when we applied Direct Maximal Center Gauge to our $SU(2)$ configurations \cite{Osaka}.
One may wonder if the center d.o.f. are responsible for some non-perturbative
features of the Yang-Mills theory, like the string tension, and not for others,
like the glueball mass. Or one may speculate that a local effective Hamiltonian
can indeed be defined after center projection, but that an imperfect 
identification of the center d.o.f. spoils this construction: this would be
one way to explain why a proper, decreasing behaviour of the
Creutz ratios is restored in the Laplacian gauge.
One bold attitude towards this problem is to make an ansatz for this
effective Hamiltonian and test its consequences \cite{Engelhardt}.
Much work remains to be done in studying the role and the interaction of
the center d.o.f.


Here we have tried to put on firmer numerical ground the first step in this 
ambitious program, namely that $Z_N$ center degrees of freedom are responsible for the
full string tension of the $SU(N)$ theory. In this process, 
we have clarified the meaning of center vortices and Abelian monopoles
as local gauge defects: each monopole is pierced by a center-vortex string.
Monopoles and anti-monopoles alternate along the center-vortex string like
beads on a necklace.
Furthermore, the eigenvectors of the covariant adjoint Laplacian, whose color
orientations are used to fix the gauge, are analogous to the adjoint Higgs
field in the Georgi-Glashow model. And our Abelian monopoles, which we 
identify via zeros of the Higgs field, are analogous to the 't Hooft-Polyakov
monopoles. Center vortices appear when a second adjoint Higgs field is
considered.

Let us speculate further on a possible scenario for the complementary role of
monopoles and center vortices.
We consider 3 distance regimes for the force between static charges:
$(i)$ At large distances $r > r_2$, center vortices are the relevant degrees
of freedom; they govern the behaviour of the Wilson loop in all group
representations, and in particular give the correct fundamental string tension
and the correct, vanishing adjoint string tension. $(ii)$ At intermediate
distances $r_1 < r < r_2$, center d.o.f. are insufficient to describe the
non-Abelian theory, and the Abelian monopoles must be considered; this
generates a linearly rising potential in the adjoint and higher representations,
with Abelian Casimir ratios (2 instead of 8/3 for adjoint $SU(2)$). $(iii)$ At short
distances $r < r_1$, the full non-Abelian nature of the color field must be
considered, to ensure recovery of the one-gluon exchange perturbative potential.
In other words, the original non-Abelian d.o.f. can be progressively decimated, 
down to Abelian then center ones, as one considers larger distances.

In this scenario, the distances $r_1$ and $r_2$ separating the three regimes are related to
the typical sizes of a monopole and a center vortex respectively:
$(i)$ an Abelian description of the monopole field is sufficient when one stays 
outside its core; and $(ii)$ at large distances, the monopole field can be 
further approximated by that of the center vortex which pierces it. 
Statement $(i)$ is well-known: the regular $SU(2)$ gauge field of 
a 't Hooft-Polyakov monopole can be gauge-transformed to a ``stringy'' gauge,
where the gauge field is singular along a Dirac string (e.g. $x=y=0, z<0$), 
and where the flux of the Dirac string is cancelled by the Higgs field. In that 
gauge, the gauge field is Abelian outside the core of the monopole. 
To justify statement $(ii)$,
one can choose a gauge where the gauge field is singular along a whole
line ($x=y=0$) rather than a half-line ($x=y=0, z<0$) \cite{Arafune}. 
In that gauge also, the
gauge field is Abelian outside the monopole core. Moreover, at large $|z|$  
it is the same as that given by a center vortex piercing the monopole along 
$x=y=0$. 

However, while the existence of a scale $r_1$ below which an Abelian description
breaks down seems inevitable, that of a separate scale $r_2 > r_1$ where
center d.o.f. ``kick in'' is less so. One may say that center vortices screen
the monopoles. 
We also find support for this scenario in theoretical studies
of the Georgi-Glashow model \cite{Ambjorn,Cornwall}, which argue that
the 't Hooft-Polyakov monopoles present in this model become irrelevant at
large distances, and that confinement is produced by center vortices instead.
Ref.\cite{Cornwall} in particular builds explicitly a classical solution which 
looks like a 't Hooft-Polyakov monopole at short distances, and a pair of center
vortices at large distances.

\vspace*{1cm}
{\bf Note Added.}
Since the completion of this paper, further work \cite{borny2} extending the study
of \cite{borny} has appeared. The results of Refs.\cite{borny,borny2} can be 
summarized as follows: in the iterative Maximal Center Gauge, a larger value
of the gauge-fixing functional yields a smaller value of the center-projected
string tension. It is then doubtful whether the latter can be equal to the 
non-Abelian string tension in the limit where the global maximum of the 
gauge-fixing functional is approached and the physical volume sent to infinity. 
The authors of \cite{borny,borny2} and those of \cite{bertle} vary in the 
interpretation of these numerical results, but concur that they
put the usefulness of the Maximal Center Gauge in jeopardy.
Our own interpretation is the following.
Because of the lattice periodic boundary conditions, every gauge defect 
(monopole in a time-slice, center vortex in a plane) must be paired with 
another opposite one. Therefore, a gauge which is sufficiently non-local may 
bring together the locations of the members of each pair, and produce no net 
defect at all after projection. This may well be what happens: in the effort to
approach the global maximum, defect pairs annihilate and the projected
string tension drops. This illustrates that the pursuit of the global maximum
is of marginal interest, whereas it is essential to pay attention to the locality of the
gauge condition (in our case, of the operator used to fix the gauge), as we have emphasized in the discussion Section V.
A similar claim that the Maximal Abelian Gauge suppresses monopoles excessively,
but that the Laplacian gauge does not, was already argued in \cite{vds}.

\vspace*{1cm}
\noindent
\underline{Acknowledgements:}
We thank C. Alexandrou, L. Cosmai, S. D\"urr, M. D'Elia, J. Fr\"ohlich, 
M. Garc\'ia P\'erez,
J. Greensite, J. Stack and T. Tomboulis for helpful discussions.

\end{document}